\documentclass{jpsj3}
\usepackage{txfonts}
\usepackage{graphicx,color,amsmath,amsxtra}
\usepackage{amssymb}
\usepackage[english]{babel}
\usepackage{amsfonts}

\allowdisplaybreaks[4]

\title{The Casimir effect as a candidate of dark energy}

\author{Jiro Matsumoto\thanks{matumoto@th.phys.nagoya-u.ac.jp, jmatsumoto@tuhep.phys.tohoku.ac.jp}}
\inst{Department of Physics, Nagoya University, 
Nagoya 464-8602, Japan\thanks{Present address: 
Department of Physics, Tohoku University, Sendai 980-8578, Japan}} 

\abst{It is known that the simply evaluated value of the zero point energy of quantum fields is extremely
deviated from the observed value of dark energy density. 
In this paper, we consider whether the Casimir 
energy, which is the zero point energy brought from boundary conditions, can cause the accelerated expansion of the Universe 
by using proper renormalization method and introducing the fermions of finite temperature living in $3+n+1$ space-time. 
We show that the zero temperature Casimir energy and the finite temperature corrections can explain 
the current accelerated expansion of the Universe. }

\kword{Dark energy, the Casimir effect, finite temperature effect.}

\begin{document}
\maketitle

\section{Introduction \label{sec1}}	

The discovery of the accelerated expansion of the Universe in late 1990's\cite{5} has brought many models of dark energy in this ten and several years. 
Because the modifications of the Einstein equation are necessary to realize the accelerating expansion when we assume that the Universe 
is isotropic and homogeneous. 
Including the cosmological constant $\Lambda$ into the Einstein equation is generally regarded as the most possible model to explain 
the accelerated expansion, because of 
its simplicity. This model, however, has a so-called fine-tuning problem\cite{7}. 
There are several ways to illustrate this problem, the fine-tuning problem 
means that the energy scale of the cosmological constant $\Lambda$ to explain the acceleration of the Universe is too small compared to the 
Planck scale $M_\mathrm{Pl}$, which is the energy scale of the gravity.  
There have been attempts to explain the accelerated expansion of the Universe not by a cosmological constant but by dynamical fields 
 without such a fine-tuning. And there are also modified gravity models to explain such an acceleration. 
As a matter of fact, fine tunings of the free parameters are usually included in these models. 

On the other hand, there are attempts to derive the observed value of the cosmological constant by the zero point energy of the quantum 
fields\cite{10,17}. 
However, there would not have been persuasive explanations so far. 
In this paper, the zero point energy generated from boundary conditions, which is called the Casimir energy, is considered. 
To be concrete, the Casimir energy of the graviton and the fermions are considered when space-time has compact extra dimensions. 
Similar investigations are seen in the literature written by B.~Greene and J.~Levin\cite{20}. 
In the literature, it is shown that the de Sitter universe is realized by introducing some bosons and some fermions live in extra dimensions. 
Whereas, there are many literatures written about the Casimir effect on various space-times without 
investigating the dynamics of the Universe. 
The features of this paper are the concrete derivation of the finite temperature Casimir effect on 
$R^3 \times (S^1)^n$ space without approximations, to show the possibility to explain dark matter by the Casimir effect, and 
to investigate the case that the extra dimensions are not static. 

In Sect.~\ref{sec2}, the finite temperature Casimir effect on $R^3 \times (S^1)^n$ space is calculated. 
The dynamics of the Universe when the Casimir effect is included is studied in Sect.~\ref{sec3} with the results of Sect.~\ref{sec2}. 
The reader not interested in the derivation of the formulas on the Casimir effect can skip Sect.~\label{sec2} and start reading from Sect.~\ref{sec3}. 
Concluding remarks are given in Sect.~\ref{sec4}. 
Units of $k_\mathrm{B} = c = \hbar = 1$ are used and the 
gravitational constant $8 \pi G$ is denoted by
${\kappa}^2 \equiv 8\pi/{M_{\mathrm{Pl}}}^2$ 
with the Planck mass of $M_{\mathrm{Pl}} = G^{-1/2} = 1.2 \times 10^{19}$GeV. 

\section{The finite temperature Casimir effect on $R^3 \times (S^1)^n$ space \label{sec2}}
In this section, we will calculate the finite temperature Casimir effect on $R^3 \times (S^1)^n$ space. 
We use the formulation and results in the literature written by S.~Bellucci and A.~A.~Saharian\cite{30}. 
The following calculations are the extensions of a part of the results in it\cite{30} 
to use them in the case of finite temperature. 
And formulas written in \textit{Integrals and Series}\cite{35} will be used in the following without provisos. 

The energy-momentum tensor of Dirac field on $R^3 \times S^n$ space is given as follows: 
\begin{eqnarray}
\langle 0 \vert T^0 _0 \vert 0 \rangle = -\frac{N}{2 d^n} \int \frac{d^3 \mbox{\boldmath$k$}}{ (2 \pi)^3}
\sum _{\textrm{\boldmath $l$}_n \in \mathbb{Z}^n} \omega_{k,l_n}, \label{3} \\
\langle 0 \vert T^i _i \vert 0 \rangle = \frac{N}{2 d^n} \int \frac{d^3 \mbox{\boldmath$k$}}{(2 \pi)^3} 
\sum _{\textrm{\boldmath $l$}_n \in \mathbb{Z}^n} \frac{k_i^2}{\omega_{k,l_n}} , \label{4}
\end{eqnarray}
where $\omega_{k,l_n}$ is defined as 
\begin{eqnarray}
\omega_{k,l_n}= \sqrt{m^2 + \mbox{\boldmath$k$}^2 
+ \sum_{i=4}^{n+3} \left ( \frac{2\pi l_i}{d} \right ) ^2 }. \label{6}
\end{eqnarray}
In Eq.~(\ref{4}), $T^i _i$ does not mean the summation with respect to $i$ and $k_i \equiv 2 \pi l_i/d$ for $i \geq 4$. 
The common periodic length $d$ are used in Eq.~({\ref{3}}) and ({\ref{4}}) because the equivalence of each extra dimension has been assumed here. 
We denote the real degrees of freedom of particles by $N$. 
These Eqs.~(\ref{3}), (\ref{4}) can be shown in the case of scalar field. 
So we assume Eqs.~(\ref{3}), (\ref{4}) are also held in the case of the graviton. 
The graviton and a sort of fermions are only considered as the degrees of freedom of particles in the following section. 

If we assume neutrinos as the fermions then 
$N = N_\nu \equiv 2^{\left [ \frac{4+n}{2} \right ]-1}$ for each generation of neutrinos and 
$N = N_g \equiv -(n+4)(n+1)/2$ for the graviton as the degrees of freedom. 
Here, we assume that the neutrinos are Dirac type. 
The negativeness of the degrees of freedom for the graviton comes from the difference of statistics. 

Then, we find the following relations from Eqs.~({\ref{3}}) and ({\ref{4}}): 
\begin{eqnarray}
p_0 = \frac{1}{3} \left \{ (n+1)\rho _0 - m \frac{\partial \rho _0}{\partial m} +d \frac{\partial \rho _0}{\partial d} \right \} , \nonumber \\
p_{b0} = - \rho _0 - \frac{d}{n} \frac{\partial \rho _0}{\partial d}, 
\label{9}
\end{eqnarray}
where $\rho _0 = \langle 0 \vert T^0 _0 \vert 0 \rangle $, $p_0 = -\sum_{i=1}^3 \langle 0 \vert T^i _i \vert 0 \rangle /3$ and 
$p_{b0} = - \sum_{i=4}^{n+3} \langle 0 \vert T^i _i \vert 0 \rangle /n$, because the signature of the metric 
is assumed to be $(+, -, -, -, -, \cdots, -)$. 
Therefore, we do not need to calculate $p_0$ and $p_{b0}$, respectively. 
Calculations of $\rho _0$ will be omitted and we only give the expression for $\rho _0$ 
because the expression of $\rho _0$ is given in former researches\cite{30, 40, 44} and 
we will conduct similar calculations to derive the finite temperature correction to the Casimir effect as follows, 
\begin{eqnarray}
\rho _0 = N \left ( \frac{m}{2 \pi} \right )^{\frac{4+n}{2}} 
\sum _{ \substack{ (l_4, l_5, \cdots , l_{n+3}) \in \mathbb{Z}^n, \\  (l_4, l_5, \cdots , l_{n+3}) \neq \textrm{\boldmath $0$} } } 
\frac{K_{\frac{4+n}{2}}\left [ md \sqrt{\sum_{i=4}^{3+n} l_i^2} \right ]}
{d^{\frac{4+n}{2}} \sqrt{\sum_{i=4}^{3+n} l_i^2}^{\frac{4+n}{2}}}. \label{12}
\end{eqnarray}
Here, $K_n(z)$ is the modified Bessel function of the second kind defined by 
\begin{eqnarray}
K_n(z)=\frac{\left (\frac{z}{2} \right )^n \Gamma \left ( \frac{1}{2} \right )}{\Gamma \left (n+ \frac{1}{2} \right )}
\int _1 ^\infty \mathrm{e}^{-zt} (t^2-1)^{\frac{2n-1}{2}} dt. 
\label{60}
\end{eqnarray}
In the case of massless particles, Eq.~(\ref{12}) is simplified to 
\begin{eqnarray}
\rho _0 = N \frac{\Gamma \left ( \frac{4+n}{2} \right )}{2 \pi ^{\frac{4+n}{2}} d^{4+n}} 
\sum _{\substack{(l_4, l_5, \cdots , l_{n+3}) \in \mathbb{Z}^n,  \\  (l_4, l_5, \cdots , l_{n+3}) \neq \textrm{\boldmath $0$} } } 
\left ( \sum_{i=4}^{3+n} l_i^2 \right ) ^{-\frac{4+n}{2}}. 
\label{15}
\end{eqnarray}
By using the Matsubara formalism and the zeta function regularization, we obtain the following expression of the 
free energy\cite{50}: 
\begin{align}
F_0 = E_0 ^{\mathrm{ren}} + \Delta _T F_0, \\
 \Delta _T F_0 \equiv k_B T \sum_J \ln \left ( 1\mp \mathrm{e}^{- \beta \omega _J} \right ), 
\label{70}
\end{align}
where $J$ indicates the eigenvalue of the momentum and $E_0 ^{\mathrm{ren}}$ is the Casimir energy of zero temperature. 
In Eq.~(\ref{70}), the above sign of $\mp$ is applied to bosons and the below sign of $\mp$ is applied to fermions. 
The expression of $E_0 ^{\mathrm{ren}}= Vd^n \rho _0$ is given in Eq.~(\ref{12}) so that we consider the density of $\Delta _T F_0$: 
\begin{align}
\frac{\Delta _T F_0}{(Vd^n)} = \frac{1}{d^n \beta} \int \frac{d^3 {\mbox{\boldmath$k$}}}{(2 \pi)^3} \sum _{(l_4, l_5, \cdots , l_{n+3}) \in \mathbb{Z}^n}
\ln \left ( 1 \mp \mathrm{e} ^{- \beta \sqrt{m^2 + {\mbox{\boldmath$k$}}^2 + (2 \pi/d)^2 \sum _{i=4}^{n+3}l_i^2 } } \right ), 
\label{80}
\end{align}
where $V$ is a volume of the three dimensional space. 
The density of the free energy is not singular in the expression of Eq.~(\ref{80}). 
However, it would include the unphysical energy density when we are confined in the compact geometry. 
Therefore, we define the density of the free energy $\Delta _T f_0$ by subtracting the zero point energy without boundary conditions from Eq.~(\ref{80}) 
as defined in the Casimir energy of zero temperature: 
\begin{align}
\Delta _T f_0 =& \frac{\vert N \vert}{d^n \beta} \int \frac{d^3 {\mbox{\boldmath$k$}}}{(2 \pi)^3} \sum _{(l_4, l_5, \cdots , l_{n+3}) \in \mathbb{Z}^n}
\ln \left ( 1 \mp \mathrm{e} ^{- \beta \sqrt{m^2 + {\mbox{\boldmath$k$}}^2 + (2 \pi/d)^2 \sum _{i=4}^{n+3}l_i^2 } } \right ) \nonumber \\
&- \frac{\vert N \vert}{\beta} \int \frac{d^{n+3} {\mbox{\boldmath$k$}}}{(2 \pi)^{n+3}} 
\ln \left ( 1 \mp \mathrm{e} ^{- \beta \sqrt{m^2 + {\mbox{\boldmath$k$}}^2 } } \right ) 
\nonumber \\
=& - \frac{\vert N \vert}{d^n \beta} \int \frac{d^3 {\mbox{\boldmath$k$}}}{(2 \pi )^3} \sum _{s=1}^\infty \frac{(\pm 1)^s}{s} 
\left \{ \sum _{(l_4, l_5, \cdots , l_{n+3}) \in \mathbb{Z}^n}
\mathrm{e} ^{- s \beta \sqrt{m^2 + {\mbox{\boldmath$k$}}^2 + (2 \pi/d)^2 \sum _{i=4}^{n+3}l_i^2 } } \right.
\nonumber \\
&-\left. d^n \int \frac{d^n {\mbox{\boldmath$k$}}'}{(2 \pi)^n} \mathrm{e}^
{ -s \beta \sqrt{m^2 + {\mbox{\boldmath$k$}}^2 +{\mbox{\boldmath$k$}}'^2 }} \right \} 
\nonumber \\
=& - \frac{\vert N \vert}{d^n \beta} \sum _{s=1}^\infty \frac{(\pm 1)^s}{s} \sum _{j=0}^{n-1} \Delta _j(s,d,\beta), \label{95} \\
&\Delta _j(s,d,\beta) \equiv d^j \int \frac{d^{3+j} {\mbox{\boldmath$k$}}}{(2 \pi)^{3+j}} 
\sum _{(l_{5+j}, \cdots , l_{n+3}) \in \mathbb{Z}^{n-j-1}} \left \{ \sum _{l_{4+j} \in \mathbb{Z}}
\mathrm{e} ^{- s \beta \sqrt{m^2 + {\mbox{\boldmath$k$}}^2 + (2 \pi/d)^2 \sum _{i=4+j}^{n+3}l_i^2 } } \right.
\nonumber \\
&-\left. d \int _{- \infty} ^{\infty} \frac{d k_{j+4}}{2 \pi} \mathrm{e}^
{ -s \beta \sqrt{m^2 + {\mbox{\boldmath$k$}}^2 +k_{j+4}^2 + (2 \pi/d)^2 \sum _{i=5+j}^{n+3}l_i^2}} \right \} . 
\label{100}
\end{align}
Eq.~(\ref{100}) can be transformed into 
\begin{align}
\Delta _j(s, d, \beta)=& 4s \beta d^{j+1} \sum _{(l_{5+j}, \cdots , l_{n+3}) \in \mathbb{Z}^{n-j-1}} \sum_{u=1}^{\infty}
\left \{ \frac{ \sqrt{ m^2  + (2 \pi /d  )^2 \sum _{i=5+j}^{n+3}l_i^2} }{2 \pi } \right \} ^{\frac{5+j}{2}} \nonumber \\
& \times \frac{K_{\frac{j+5}{2}} \left [ \sqrt{m^2  + (2 \pi /d  )^2 \sum _{i=5+j}^{n+3}l_i^2 } \sqrt{ s^2 \beta ^2 + u^2 d^2}   \right ] }
{ \left [  \sqrt{ s^2 \beta ^2 + u^2 d^2}  \right ]^{\frac{j+5}{2}} }. 
\label{165}
\end{align} 
The derivation of Eq.~(\ref{165}) is written down in Appendix \ref{app}. 
Substituting Eq.~(\ref{165}) into Eq.~(\ref{95}), we obtain 
\begin{align}
\Delta _T f_0 =& - 4 \vert N \vert d^{j-n+1} \sum _{s=1}^\infty (\pm 1)^s \sum _{j=0}^{n-1} 
\sum _{(l_{5+j}, \cdots , l_{n+3}) \in \mathbb{Z}^{n-j-1}} \sum_{u=1}^{\infty} 
\left \{ \frac{ \sqrt{ m^2  + (2 \pi /d  )^2 \sum _{i=5+j}^{n+3}l_i^2} }{2 \pi } \right \} ^{\frac{5+j}{2}} \nonumber \\
& \times \frac{K_{\frac{5+j}{2}} \left [ \sqrt{m^2  + (2 \pi /d  )^2 \sum _{i=5+j}^{n+3}l_i^2 } \sqrt{ s^2 \beta ^2 + u^2 d^2}   \right ] }
{ \left [  \sqrt{ s^2 \beta ^2 + u^2 d^2}  \right ]^{\frac{5+j}{2}} }. 
\label{170}
\end{align} 
Eq.~(\ref{170}) can be rewritten in more simple form as: 
\begin{align}
\Delta _T f_0 = - \frac{2 \vert N \vert m^{4+n}}{(2 \pi)^{\frac{4+n}{2}} } \sum _{s=1} ^\infty (\pm 1)^s 
\sum _{(l_{4}, \cdots , l_{n+3}) \in \mathbb{Z}^n} 
\frac{K_{\frac{4+n}{2}}\left [ m \sqrt{\beta ^2 s^2 + d^2 \sum_{i=4}^{3+n} l_i^2} \right ]}
{m^{\frac{4+n}{2}}\sqrt{\beta ^2 s^2 + d^2 \sum_{i=4}^{3+n} l_i^2}^{\frac{4+n}{2}}}. 
\label{180}
\end{align} 
The way of this transformation is identical with that in the appendix of S.~Bellucci and A.~A.~Saharian\cite{30}. 
To explain the way of the transformation briefly, making $\Delta _j (s, \beta , d)$ from Eq.~(\ref{180}) 
and comparing it to Eq.~(\ref{165}) gives the equivalence of Eqs.~(\ref{170}) and (\ref{180}). 
From Eqs.~(\ref{12}) and (\ref{180}), we obtain the finite temperature Casimir energy density $\rho _\mathrm{casimir}$ by considering the 
definition of the free energy as 
\begin{align}
\rho _\mathrm{casimir} = -\frac{\partial}{\partial \beta}\left \{ \beta \frac{\vert N \vert m^{4+n}}{(2 \pi)^{\frac{4+n}{2}}} 
\sum _{ \substack{s \in \mathbb{Z}, \\ (l_{4}, \cdots , l_{n+3}) \in \mathbb{Z}^n, \\ (s, l_{4}, \cdots , l_{n+3}) \neq \textrm{\boldmath $0$} } }
(1-2 \delta _{s0}) ^\alpha (-1)^{\alpha s} 
\frac{K_{\frac{4+n}{2}}\left [ m \sqrt{\beta ^2 s^2 + d^2 \sum_{i=4}^{3+n} l_i^2} \right ]}
{m^{\frac{4+n}{2}}\sqrt{\beta ^2 s^2 + d^2 \sum_{i=4}^{3+n} l_i^2}^{\frac{4+n}{2}}} \right \}, 
\label{190}
\end{align}
where the notation $(-1)^\alpha = 1$ for bosons and $(-1)^\alpha = -1$ for fermions is used. 
\section{Cosmic acceleration caused by the Casimir effect \label{sec3}}
In this section, we apply the result in the last section to the space-time of the expanding universe. 
We, then, have to care about the time-varying metric and the existence of the baryons and the photons, which do not propagate into the extra dimensions. 
However, we can neglect such properties and apply the result by replacing the periodic length $d$ and the thermodynamic beta $\beta$ 
with $d(t)$ and $\beta (t)$, respectively. 
Because the time-varying scale of the Universe $H(t)\sim 10^{-33}$eV is much less than 
the energy scale of dark energy $\sim 10^{-3}$eV, which is brought by the Casimir effect. 
And the energy density of the baryons and the photons are also small compared to 
that of dark energy and dark matter. 
Therefore, the inhomogeneity for the direction of the extra dimension would be negligible when 
we consider the Casimir effect as a root of dark matter and dark energy. 
\subsection{Friedmann-Lemaitre equations}
First, we assume Friedmann-Lemaitre-Robertson-Walker (FLRW) metric of the flat space, 
\begin{align}
ds^2 = -dt^2 + a^2(t) \delta_{ij}dx^i dx^j + b^2(t) \delta_{\diamondsuit \heartsuit} dx^\diamondsuit dx^\heartsuit, 
\label{c10}
\end{align}
where $i,j=1,2,3$ and $\diamondsuit , \heartsuit = 4, 5, \cdots, n+3$. 
And $\mu , \nu = 0,1,2, \cdots, n+3$ will be used. 
Then, we can write down Friedmann-Lemaitre (FL) equations as follows:  
\begin{align}
3H^2+3nHH_b+\frac{1}{2}n(n-1)H_b^2=\kappa ^2 d_0^n \rho, \label{c20} \\
-3H^2-2\dot H -2nHH_b -\frac{1}{2}n(n+1)H_b ^2 -n\dot H_b = \kappa ^2 d_0^n p, \label{c30} \\
-6H^2-3 \dot H +3(1-n)HH_b -\frac{1}{2}n(n-1)H_b ^2 +(1-n)\dot H_b = \kappa ^2 d_0^n p_b. 
\label{c40}
\end{align} 
Here, $d(t)=d_0 b(t)$ is the periodic length of the compact dimensions and $d_0$ is the current value of it. 
The Hubble rate of the three dimensions is defined by $H(t) \equiv \dot a(t)/ a(t) $ and the expansion rate of extra dimensions is defined by 
$H_b(t) \equiv \dot b(t)/ b(t)$. $\rho$, $p$ and $p_b$ are defined as $\rho = - T^0_0$, $p = T^i_i /3$ and $p_b = T^\diamondsuit _\diamondsuit /n$, 
respectively. 
By eliminating the terms $\dot H(t)$ and $H^2 (t)$ from Eq.~(\ref{c40}), the equation of motion for the size of 
extra dimensions is derived: 
\begin{align}
\frac{1}{a^3b^n} \frac{d}{dt} \left ( a^3 b^n H_b \right ) = \frac{\kappa ^2}{n+2} d_0^n (\rho + 2p_b -3p). 
\label{c50}
\end{align} 
On the other hand, we obtain the equation of continuity $\nabla _\mu T^{\mu 0}=0$ in the following form, 
\begin{align}
\dot \rho + 3H(\rho + p) + nH_b (\rho +p_b)=0. 
\label{c60}
\end{align} 
The Casimir energy density, pressure and the pressure in the extra dimensions of zero temperature are given by 
Eqs.~(\ref{9}) and (\ref{12}). Substituting Eq.~(\ref{12}) into Eq.~(\ref{9}) for $p_0$ yields the relation $p_0 = -\rho_0$, correctly. 
Whereas, from the definition of the free energy, we obtain the following expressions of the total pressures: 
\begin{align}
p_\mathrm{casimir} &= \frac{\vert N \vert m^{4+n}}{(2 \pi)^{\frac{4+n}{2}}} 
\sum _{ \substack{s \in \mathbb{Z}, \\ (l_{4}, \cdots , l_{n+3}) \in \mathbb{Z}^n, \\ (s, l_{4}, \cdots , l_{n+3}) \neq \textrm{\boldmath $0$} } }
(1-2 \delta _{s0}) ^\alpha (-1)^{\alpha s} 
\frac{K_{\frac{4+n}{2}}\left [ m \sqrt{\beta ^2 s^2 + d^2 \sum_{i=4}^{3+n} l_i^2} \right ]}
{m^{\frac{4+n}{2}}\sqrt{\beta ^2 s^2 + d^2 \sum_{i=4}^{3+n} l_i^2}^{\frac{4+n}{2}}} , \label{c70} \\
p_{b,\mathrm{casimir}} &=  \frac{\partial}{\partial d^n}\left \{ d^n \frac{\vert N \vert m^{4+n}}{(2 \pi)^{\frac{4+n}{2}}} 
\sum _{ \substack{s \in \mathbb{Z}, \\ (l_{4}, \cdots , l_{n+3}) \in \mathbb{Z}^n, \\ (s, l_{4}, \cdots , l_{n+3}) \neq \textrm{\boldmath $0$} } }
(1-2 \delta _{s0}) ^\alpha (-1)^{\alpha s} 
\frac{K_{\frac{4+n}{2}}\left [ m \sqrt{\beta ^2 s^2 + d^2 \sum_{i=4}^{3+n} l_i^2} \right ]}
{m^{\frac{4+n}{2}}\sqrt{\beta ^2 s^2 + d^2 \sum_{i=4}^{3+n} l_i^2}^{\frac{4+n}{2}}} \right \}. 
\label{c80}
\end{align}
These Eqs.~(\ref{c70}) and (\ref{c80}) are consistent with equations in (\ref{9}). 

Next question is the $a(t)$ and $b(t)$ dependence of the temperature. 
From the definition of the free energy, we have the entropy: 
\begin{align}
S_\mathrm{casimir} \propto - \frac{ \partial (\rho _\mathrm{casimir} a^3 b^n)}{\partial T} \propto a^3 b^n \beta ^2 \frac{\partial^2 ( \beta \Delta _T f_0)}{\partial \beta^2}. 
\end{align} 
The adiabatic expansion of the Universe implies, 
\begin{align}
dS_\mathrm{casimir} = 0.
\label{ec}
\end{align}
The condition (\ref{ec}) is equivalent to the equation of continuity (\ref{c60}). 
Thus, if the term proportional to $\beta  ^{-4-n}(t)$ 
is a dominant component of $\Delta _T f_0$, then we have, 
\begin{align}
\beta (t) \propto a^{\frac{3}{3+n}}(t) b^{\frac{n}{3+n}} (t). 
\label{c100}
\end{align} 
The entropy conservation for the photon and the particles which cause the Casimir effect induce the following 
relation between the current background temperature of its particles $T_0$ and that of the photon $T_{\gamma 0}$ as: 
\begin{align}
T_0 = \left ( \frac{2}{\mathcal{N}_\mathrm{dec}} \right ) ^{\frac{1}{3}}
 \left ( \frac{a_0 b_\mathrm{dec}}{b_0 a_\mathrm{dec}} \right ) ^{\frac{n}{3+n}} T_{\gamma 0},  \label{c150}
\end{align}
where $T_{\gamma 0}=2.725$K, $\mathcal{N}_\mathrm{dec}$ is the number of particle types with an extra factor of $7/8$ for fermions 
in equilibrium with the photon at the time of decoupling. 
We have only considered the leading terms of the entropy, therefore there appear correction terms. 
We can find that $T_0$ of this case can be larger than $2.725$K if $a_0/a_\mathrm{dec} > b_0/b_\mathrm{dec}$. 
It is necessary to be careful to the fact that the temperature given in Eq.~(\ref{c150}) is not the temperature of three dimensional spaces but 
the effective temperature of $3+n$ dimensional spaces. 
\subsection{Expansion history of the Universe}

There are two scenarios to explain the current accelerating expansion of the Universe in this model. 
The first one is $\dot b(t) =0$ and the second one is $\dot b(t) <0$. 

By expanding the expression in Eq.~(\ref{190}), we can see that there are terms proportional to $b^{-n-4}(t)$ and 
$\beta ^{-n-4}(t)$, which are dominant terms in Eq.~(\ref{190}). 
The former one comes from the zero temperature Casimir energy, and the latter one is a contribution from 
the finite temperature corrections. 
In the case of $\dot b(t) = 0$, the terms proportional to $b^{-n-4}(t)$ behave as dark energy 
because they are just like a cosmological constant. 
If the term proportional to $\beta  ^{-4-n}(t)$ is a dominant component of $\Delta _T f_0$, then Eq.~(\ref{c100}) is 
satisfied. 
Therefore, the terms proportional to $\beta  ^{-4-n}(t)$, then, behave as dark matter. 
In the following, We investigate whether or not this scenario is realized in the case of $n=1$. 
The Casimir energy density is given as: 
\begin{align}
\rho _\mathrm{casimir} =& - \frac{1}{4 \pi ^2 } \sum _{ \substack{s, l_4 \in \mathbb{Z}, \\ (s,l_4) \neq (0,0) } } \frac{\partial}{\partial \beta}
\left [ \frac{15}{2} \frac{\beta}{(s^2 \beta ^2 + l_4^2 d^2)^{\frac{5}{2}} } \right. \nonumber \\ 
&+  \frac{N_{\psi}}{2} \beta (1- 2 \delta _{s0}) (-1)^s \frac{m ^2 \mathrm{e}^{-m \sqrt{s^2 \beta ^2 + l_4 ^2 d^2} } }
{(s^2 \beta ^2 + l_4 ^2 d^2)^{\frac{3}{2}} } \nonumber \\
&\times \left. \left \{ 1+ \frac{3}{m \sqrt{s^2 \beta ^2 + l_4 ^2 d^2}} + \frac{3}{m ^2 \left (s^2 \beta ^2 + l_4 ^2 d^2 \right ) } \right \} \right ]. 
\label{c110}
\end{align}
The first line of Eq.~(\ref{c110}) are contributions from the graviton, and the second line and the third line of Eq.~({\ref{c110}}) 
are contributions from the massive fermions, respectively. 
The leading terms of the energy density are given by 
\begin{align}
\rho _\mathrm{casimir} \sim &- \frac{15}{8 \pi ^2} \left ( \frac{1}{d^5} - \frac{4}{\beta ^5} \right ) \nonumber \\
&+ \frac{N_{\psi}}{8 \pi ^2} \left \{ \mathrm{e}^{- m d} \left ( \frac{3}{d^5} + \frac{3m}{d^4} + \frac{m^2}{d^3} \right )
- \mathrm{e}^{- m \beta} \left ( \frac{12}{\beta ^5} + \frac{12m}{ \beta ^4} + \frac{5 m^2}{\beta ^3} + \frac{m^3}{\beta ^2} \right ) \right \} .
\label{c120}
\end{align}
At the same time, the balancing condition, $\rho + 2p_b -3p$, is rewritten by 
\begin{align}
\rho + 2p_b -3p 
\sim &  - \frac{15}{8 \pi ^2} \left ( \frac{12}{d^5} - \frac{3}{\beta ^5} \right ) 
+ \frac{N_{\psi}}{8 \pi ^2} \left \{ \mathrm{e}^{- m d} 
 \left ( \frac{36}{d^5} + \frac{36m}{d^4} + \frac{14m^2}{d^3} + \frac{2m^3}{d^2} \right ) \right. \nonumber \\ 
&- \left. \mathrm{e}^{- m \beta} 
\left ( \frac{9}{\beta ^5} + \frac{9m}{\beta ^4} + \frac{4m^2}{\beta ^3} + \frac{m^3}{\beta ^2} \right ) \right \} 
.\label{c130}
\end{align}
We find that if $d \gg \beta$ then $\rho + 2p_b -3p$ can be zero when $\rho _\mathrm{casimir} > 0$, else if $d \ll \beta$ then 
$\rho + 2p_b -3p$ can be zero when $\rho _\mathrm{casimir} < 0$ by combining Eqs.~(\ref{c120}) and (\ref{c130}). 
Therefore, the Casimir energy from the graviton and a sort of fermions cannot be a candidate of dark energy if the finite temperature 
corrections are negligible. 
Even if a sort of bosons are introduced instead of fermions, conditions $\rho _\mathrm{casimir} > 0$ 
and $\dot b(t) = 0$ cannot be satisfied. 
Because of this condition, several kinds of new particles are introduced to realize the de Sitter universe by the zero temperature Casimir effect 
in \cite{20}. 
On the other hand, if we consider the finite temperature effect then the conditions $\dot b(t) \approx 0$ and $\rho _\mathrm{casimir} > 0$ 
can be realized by introducing a sort of fermions and choosing the appropriate values for parameters. 
Moreover, this finite temperature corrections can be a candidate of dark matter. 
We need to be careful that the condition $\dot b(t) = 0$ is not exactly satisfied, because $\beta(t)$ depends on $a(t)$. 

Next, let us consider the case of $\dot b(t) < 0$. 
The terms proportional to $b^{-n-4}(t)$, then, contribute to dark energy because the absolute value of them grows with time evolution. 
In the case of usual $3+1$ dimensional universe, such a growth of the energy density is realized by phantom, 
which is a fluid satisfies $p/\rho = w < -1$. 
There are the other contributions for dark energy in the case of $\dot b(t) < 0$. 
The usual $3+1$ dimensional Friedmann equations do not contain the terms proportional to $H_b$ or $\dot H_b$ in Eqs.~(\ref{c20}) and (\ref{c30}). 
So these terms are treated as dark matter and dark energy. 
In particular, $3n H H_b$ in Eq.~(\ref{c20}) and $-2nHH_b$ in Eq.~(\ref{c30}) can be looked upon dark energy, because $w=-2/3 < -1/3$. 

In the following, we utilize numerical calculations for the analysis of the expansion history of the Universe. 
The equations adopted for numerical calculations are Eq.~(\ref{c20}), Eq.~(\ref{c60}) and the equation, 
\begin{equation}
\frac{1}{a^3b^n} \frac{d}{dt} \left ( a^3 b^n H \right ) = \frac{\kappa ^2}{n+2} d_0^n \left \{ \rho + (n-1)p -n p_b \right \},  
\label{c140}
\end{equation}
which is obtained from Eqs.~(\ref{c20})-(\ref{c40}). 
It is necessary to be careful that there are two solutions in the case of $n \neq 1$ because 
Eq.~(\ref{c20}) has two algebraic solutions for $H_b$. 

\begin{figure}  
\begin{minipage}[t]{0.5\columnwidth}  
\begin{center}  
\includegraphics[clip, width=0.97\columnwidth]{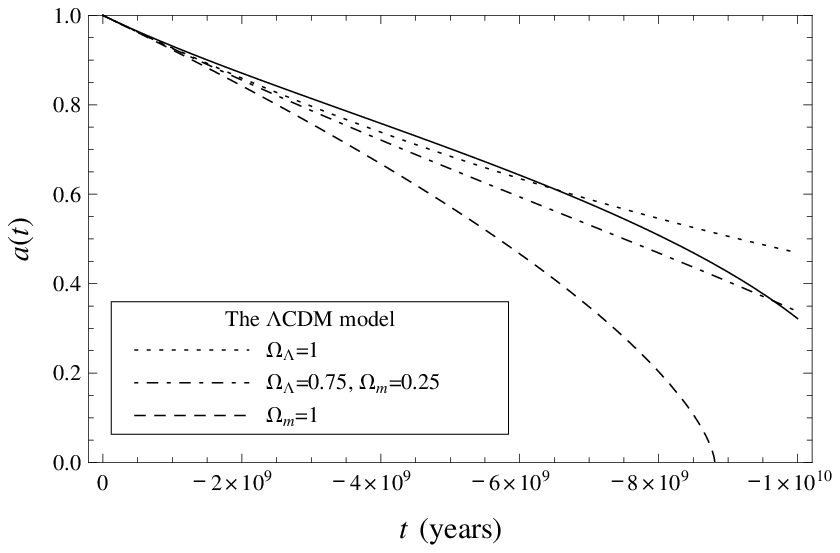}  
\end{center}  
\end{minipage}%
\begin{minipage}[t]{0.5\columnwidth}  
\begin{center}  
\includegraphics[clip, width=0.97\columnwidth]{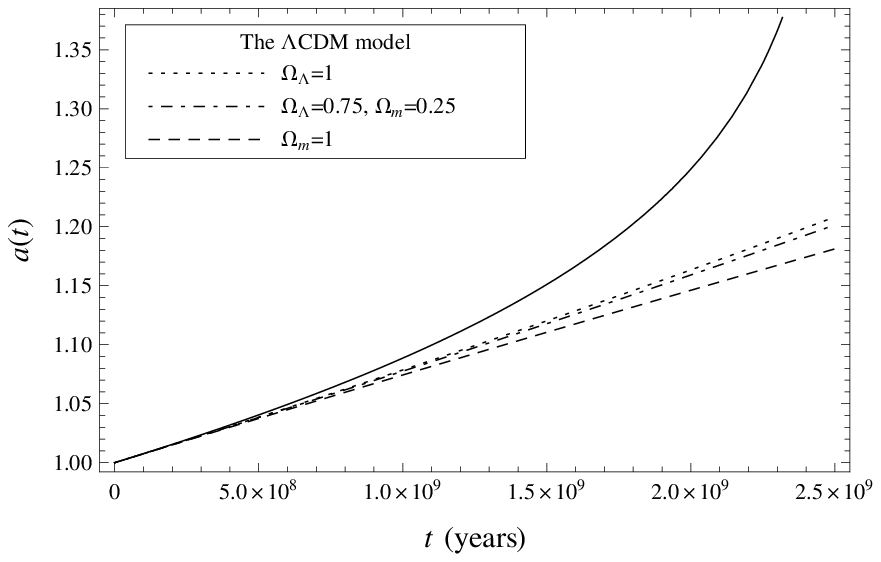}  
\end{center}  
\end{minipage}  
\caption{The solid line indicate the time evolution of the scale factor in this model, 
and the time evolutions of the scale factor in the $\Lambda$CDM model are expressed as the broken lines. 
The parameters of the model are fixed as $n=1$, $d_0=60 \: \mu$m, $T _0 = 17$ K, and $H_0=74$ km s$^{-1}$ Mpc$^{-1}$. 
Then, the relation $H_{b0} \approx -1.87 H_0$ is derived from Eq.~(\ref{c20}). 
The scale factors $a(t)$ and $b(t)$ are normalized as $a_0=b_0=1$. }  
\label{f1}
\end{figure}

\begin{figure}  
\begin{minipage}[t]{0.5\columnwidth}  
\begin{center}  
\includegraphics[clip, width=0.97\columnwidth]{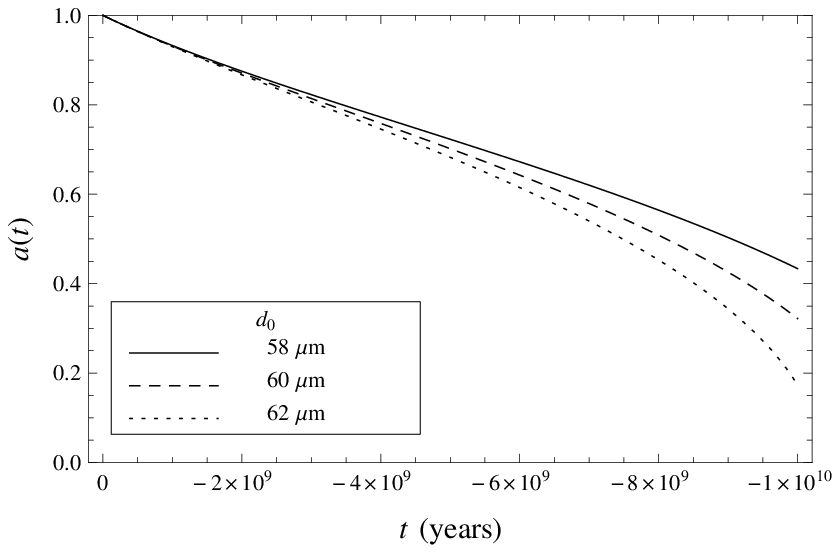}  
\end{center}  
\end{minipage}%
\begin{minipage}[t]{0.5\columnwidth}  
\begin{center}  
\includegraphics[clip, width=0.97\columnwidth]{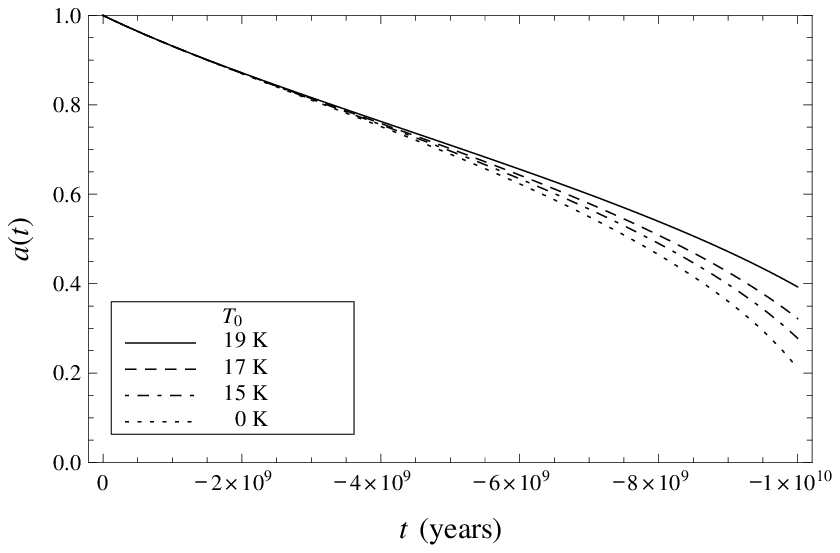}  
\end{center}
\end{minipage}  
\caption{The dependence of the time evolution of the scale factor on 
the value of the parameter $d_0$ or $T_0$. }
\label{f2}
\end{figure}

An example is shown in Fig.~\ref{f1}.  
The horizontal axis represents the time $t$ when the current value of it is zero, and the vertical axis represents the scale farctor 
$a(t)$. 
This figure is plotted by considering only the Casimir effect from the graviton of finite temperature 
without introducing the fermions which can go through the extra dimensions. 
The left figure of Fig.~\ref{f1} indicates that the expansion history of the Universe similar to that of the $\Lambda$CDM model can 
be realized. 
On the other hand, the right figure of Fig.~\ref{f1} expresses the future evolution of the scale factor. 
It indicates that a phantom-like extreme expansion of the Universe will be occur rather than the exponential expansion. 
The case of $n=1$ is considered in the above, whereas, we can also describe a $\Lambda$CDM-like evolution of the Universe in the higher dimensional model. 
The value of the parameters, $d_0$ and $T_0$, however, are changed as $d_0=90 \: \mu$m, $T _0 = 15$ K in the case of $n=2$, 
$d_0=115 \: \mu$m, $T _0 = 14$ K in the case of $n=3$, and so on. 
Such a tendency that the periodic length becomes longer and the temperature falls down, 
is come from the increment of $l$ summations in Eqs.~(\ref{190}), (\ref{c70}) and (\ref{c80}) with the augmentation of the dimension, 
to keep the total amount of the energy density and the pressures steady. 
In Fig.~\ref{f2}, the behaviors of the scale factor $a(t)$ are described when the current periodic length of the 
extra dimension $d_0$ or the current temperature of the background graviton $T_0$ is changed. 
There are a little difference in the behavior of the scale factor for the change of $d_0$ or $T_0$. 

\section{Conclusions \label{sec4}}

It has been investigated whether or not the Casimir effect from a sort of fermions and the graviton can explain dark energy if they are the only 
particles which can go through the compact extra dimensions. 
In Sect.~\ref{sec2}, the finite temperature Casimir effect has been calculated on $R^3 \times (S^1)^n$ space. 
In Sect.~\ref{sec3}, the result obtained in Sect.~\ref{sec2} are applied to the Friedmann universe, and the expansion history of the Universe has 
been investigated by using a case analysis when the Casimir energy behaves as dark energy. 
The first case is that the extra dimensions are stabilized and the accelerated expansion like a de Sitter universe 
is realized. 
In the second case, the contraction of the extra dimensions, which is equivalent to $H_b < 0$, 
plays the role of dark energy and dark matter effectively. 
The first case had been investigated in \cite{20} when the finite temperature effect is negligible. 
It is, however, found that the finite temperature corrections can simplify the configurations of the model and 
can be a candidate of dark matter. 
The numerical calculations have been carried out in the second case. 
As a result, it has been shown that the almost same expansion history of the Universe compared to the $\Lambda$CDM model 
can be realized by the finite temperature Casimir effect of the graviton without introducing fermions which can go through the extra dimensions. 
The feature of this model is that the current expansion can be described only by several physical parameters without 
introducing new fields. 
In the case of $n=1$, the values of the parameters are determined as $d_0=60 \: \mu$m and $T _0 = 17$ K. 

\begin{acknowledgment}
The author would thank H.~Yoda for daily discussions and thank S.~Nojiri for corrections and advices for the first version of the article. 
The author would also like to acknowledge the help from A.~Liliades, A.~Cerny and M.~Nami. 
This investigation is supported by Grant-in-Aid for JSPS Fellows \#461001209. 
\end{acknowledgment}

\appendix

\section{\label{app}}
$\Delta _j(s, d, \beta)$ defined in Eq.~(\ref{100}) can be arranged by using the Abel-Plana formula, 
\begin{align}
\sum_{n=0}^{\infty} F \left ( n\right) - \int_0 ^\infty dt F(t) 
= \frac{1}{2} F(0) + i \int _0 ^\infty \frac{dt}{\mathrm{e}^{2 \pi t}-1} \left [ F(it) - F(-it) \right ], 
\label{110}
\end{align}
as follows, 
\begin{align}
\Delta _j(s, d, \beta)=& d^j \int \frac{d^{3+j} {\mbox{\boldmath$k$}}}{(2 \pi)^{3+j}} 
\sum _{(l_{5+j}, \cdots , l_{n+3}) \in \mathbb{Z}^{n-j-1}} \frac{2d}{\pi} \sqrt{m^2+{\mbox{\boldmath$k$}}^2 
+\left ( \frac{2 \pi}{d} \right )^2 \sum_{i=5+j}^{n+3}l_i^2}
\nonumber \\
& \times 
\int _1 ^{\infty} \frac{dt}{\mathrm{e}^{t d \sqrt{m^2 + {\mbox{\boldmath$k$}}^2 + (2 \pi/d)^2 \sum _{i=5+j}^{n+3}l_i^2}  } -1} 
\nonumber \\
& \times \sin \left [ s \beta \sqrt{m^2 + {\mbox{\boldmath$k$}}^2 + \left ( \frac{2 \pi}{d} \right )^2 \sum _{i=5+j}^{n+3}l_i^2} \sqrt{t^2-1} \right ], 
\label{120}
\end{align}
where the formula, 
\begin{align}
G_A^{(\alpha)}(it)-G_A^{(\alpha)}(-it)= 2i \mathrm{e}^{\alpha \ln (t^2 - A^2)} \sin [\pi \alpha] \theta (t-A), \nonumber \\
G_A^{(\alpha)} (z) \equiv \exp [\alpha \ln (A^2+z^2)], 
\label{130}
\end{align}
has been used for $A= \frac{d}{2 \pi} \sqrt{m^2 + {\mbox{\boldmath$k$}}^2 + (2 \pi/d)^2 \sum _{i=5+j}^{n+3}l_i^2}$. 
The factor $1/(\mathrm{e}^x-1)$ is expanded with respect to $\mathrm{e}^{-x}$ as follows to integrate with respect to $k$ and $t$: 
\begin{align}
\Delta _j(s, d, \beta)=& d^j \int \frac{d^{3+j} {\mbox{\boldmath$k$}}}{(2 \pi)^{3+j}} 
\sum _{(l_{5+j}, \cdots , l_{n+3}) \in \mathbb{Z}^{n-j-1}} \frac{2d}{ \pi} \sqrt{m^2+{\mbox{\boldmath$k$}}^2 
+\left ( \frac{2 \pi}{d} \right )^2 \sum_{i=5+j}^{n+3}l_i^2}
\nonumber \\
& \times 
\int _1 ^{\infty} dt \sum_{u=1} ^{\infty} \mathrm{e}^{-u t d \sqrt{m^2 + {\mbox{\boldmath$k$}}^2 + (2 \pi/d)^2 \sum _{i=5+j}^{n+3}l_i^2}  } \nonumber \\
& \times \sin \left [ s \beta \sqrt{m^2 + {\mbox{\boldmath$k$}}^2 + \left ( \frac{2 \pi}{d} \right )^2 \sum _{i=5+j}^{n+3}l_i^2} \sqrt{t^2-1} \right ]. 
\label{140}
\end{align}
Furthermore, the expansion of sinusoidal function with respect to its argument enables us to examine $t$ integration: 
\begin{align}
\Delta _j(s, d, \beta)=& -d^j \int \frac{d^{3+j} {\mbox{\boldmath$k$}}}{(2 \pi)^{3+j}} 
\sum _{(l_{5+j}, \cdots , l_{n+3}) \in \mathbb{Z}^{n-j-1}} \sum_{u=1}^{\infty} \sum_{v=1}^{\infty} \nonumber \\
& \times \frac{2d}{\pi ^{3/2}s \beta} \left ( -\frac{2 s^2 \beta^2 }{ud} \right )^v \frac{\mathrm{\Gamma} (v+1/2)}{(2v-1)!} 
\left \{ m^2 + {\mbox{\boldmath$k$}}^2 + \left ( \frac{2 \pi}{d} \right )^2 \sum _{i=5+j}^{n+3}l_i^2 \right \} ^{\frac{v}{2}} \nonumber \\
& \times K_v \left [ud \sqrt{ m^2 + {\mbox{\boldmath$k$}}^2 + \left ( \frac{2 \pi}{d} \right )^2 \sum _{i=5+j}^{n+3}l_i^2} \right ], 
\label{150}
\end{align}
where $K_\nu (z)$ is the function called the modified Bessel function or Macdonald function defined in Eq.~(\ref{60}). 
By transforming the Cartesian coordinate into the spherical coordinate in $k$ integration gives, 
\begin{align}
\Delta _j(s, d, \beta)=& -d^j \sum _{(l_{5+j}, \cdots , l_{n+3}) \in \mathbb{Z}^{n-j-1}} \sum_{u=1}^{\infty} \sum_{v=1}^{\infty} 
\frac{2d}{\pi ^{3/2}s \beta} 
\left \{ \frac{ \sqrt{ m^2  + (2 \pi /d  )^2 \sum _{i=5+j}^{n+3}l_i^2} }{2 \pi u d} \right \} ^{\frac{3+j}{2}} \nonumber \\
& \times \left ( -\frac{2 s^2 \beta^2 }{ud} \right )^v \frac{\mathrm{\Gamma} (v+1/2)}{(2v-1)!} 
\left \{ m^2 + \left ( \frac{2 \pi}{d} \right )^2 \sum _{i=5+j}^{n+3}l_i^2 \right \} ^{\frac{v}{2}} \nonumber \\
& \times K_{v+\frac{j+3}{2}} \left [ud \sqrt{ m^2 + \left ( \frac{2 \pi}{d} \right )^2 \sum _{i=5+j}^{n+3}l_i^2} \right ] \nonumber \\
=&  -d^j \sum _{(l_{5+j}, \cdots , l_{n+3}) \in \mathbb{Z}^{n-j-1}} \sum_{u=1}^{\infty}
\frac{2d}{\pi ^{3/2}s \beta} 
\left \{ \frac{ \sqrt{ m^2  + (2 \pi /d  )^2 \sum _{i=5+j}^{n+3}l_i^2} }{2 \pi u d} \right \} ^{\frac{3+j}{2}} \nonumber \\
& \times \frac{\partial}{\partial \ln s} \sum_{v=1}^{\infty} \pi ^{\frac{1}{2}} (ud)^{\frac{j+3}{2}}
 \frac{1}{v!}  \left ( s^2 \beta ^2  \right )^v
\left \{ m^2 + \left ( \frac{2 \pi}{d} \right )^2 \sum _{i=5+j}^{n+3}l_i^2 \right \} ^{\frac{v}{2}+\frac{j+3}{4}} \nonumber \\
& \times \frac{d^v}{d^v \zeta} \left \{ \zeta ^{- \frac{j+3}{4}} K_{\frac{j+3}{2}}(\zeta ^{\frac{1}{2}}) \right \} \bigg \vert 
_{\zeta = u^2 d^2 \left \{ m^2 + \left ( \frac{2 \pi}{d} \right )^2 \sum _{i=5+j}^{n+3}l_i^2 \right \} } \nonumber \\
=&  d^j \sum _{(l_{5+j}, \cdots , l_{n+3}) \in \mathbb{Z}^{n-j-1}} \sum_{u=1}^{\infty}
\frac{2d}{\pi ^{3/2}s \beta} 
\left \{ \frac{ \sqrt{ m^2  + (2 \pi /d  )^2 \sum _{i=5+j}^{n+3}l_i^2} }{2 \pi u d} \right \} ^{\frac{3+j}{2}} \nonumber \\
& \times \pi ^{\frac{1}{2}} (ud)^{\frac{j+3}{2}} \left \{ m^2 + \left ( \frac{2 \pi}{d} \right )^2 \sum _{i=5+j}^{n+3}l_i^2 \right \} ^{\frac{j+3}{4}} 
\nonumber \\
& \times \frac{\partial}{\partial \ln s} \left \{ 
\frac{K_{\frac{j+3}{2}} \left [ \sqrt{m^2  + (2 \pi /d  )^2 \sum _{i=5+j}^{n+3}l_i^2 } \sqrt{ s^2 \beta ^2 + u^2 d^2}   \right ] }
{ \left [ \sqrt{m^2  + (2 \pi /d  )^2 \sum _{i=5+j}^{n+3}l_i^2 } \sqrt{ s^2 \beta ^2 + u^2 d^2}  \right ]^{\frac{j+3}{2}} } \right \} \nonumber \\
=&  4s \beta d^{j+1} \sum _{(l_{5+j}, \cdots , l_{n+3}) \in \mathbb{Z}^{n-j-1}} \sum_{u=1}^{\infty}
\left \{ \frac{ \sqrt{ m^2  + (2 \pi /d  )^2 \sum _{i=5+j}^{n+3}l_i^2} }{2 \pi } \right \} ^{\frac{5+j}{2}} \nonumber \\
& \times \frac{K_{\frac{j+5}{2}} \left [ \sqrt{m^2  + (2 \pi /d  )^2 \sum _{i=5+j}^{n+3}l_i^2 } \sqrt{ s^2 \beta ^2 + u^2 d^2}   \right ] }
{ \left [  \sqrt{ s^2 \beta ^2 + u^2 d^2}  \right ]^{\frac{j+5}{2}} }. 
\label{160}
\end{align}

\end{document}